\documentclass{IEEEtran}
\usepackage{color}

\usepackage{threeparttable}
\usepackage{amsthm}

\usepackage{cite}

\ifCLASSINFOpdf
\else
   \usepackage[dvips]{graphicx}
   \graphicspath{{../eps/}}
   \DeclareGraphicsExtensions{.eps}
\fi

\usepackage[cmex10]{amsmath}
\usepackage{empheq}

\usepackage{mdwmath}
\usepackage{mdwtab}

\begin{document}
%
\title{Outage Analysis and Beamwidth Optimization for Positioning-Assisted Beamforming}

\author{Bingcheng Zhu,~\IEEEmembership{Senior Member,~IEEE},
        Zaichen Zhang,~\IEEEmembership{Senior Member,~IEEE},
        Julian Cheng,~\IEEEmembership{Senior Member,~IEEE}
\thanks{Bingcheng Zhu and Zaichen Zhang are with School of Information Science and Engineering, Southeast University, Nanjing, Jiangsu, China (e-mail: \{zbc, zczhang\}@seu.edu.cn). The corresponding author of this work is Zaichen Zhang.}
\thanks{Julian Cheng is with School of Engineering,  The University of British Columbia, Kelowna, BC, Canada (e-mail: {julian.cheng@ubc.ca})}
}


%


\maketitle

\begin{abstract}
Beamforming based on channel estimation can be computationally intensive and inaccurate when the antenna array is large, while positioning information can reduce the complexity. In this work, we study the outage probability of positioning-assisted beamforming systems. Closed-form outage probability bounds are derived by considering positioning error, link distance and beamwidth. Based on the analytical result, we show that the beamwidth should be optimized with respect to the link distance and the transmit power, and such optimization significantly suppresses the outage probability.
\end{abstract}


%
\IEEEpeerreviewmaketitle

\vspace{-5 pt}
\section{Introduction}
Massive multiple-input multiple output (MIMO) technique significantly narrows radio-frequency beams, enabling more {{efficient}} power assignment to receivers. However, implementation of massive MIMO systems still {{has}} some challenges. For example, {{when the number of antennas is large, the radio-frequency (RF) chains and signal processing for channel estimation and beamforming are expensive and power-hungry} \cite{HybridBeamforming5G_Magz,HybridBeamforming_Molisch}}. Another important issue is that the feedback overhead surges with the number of antenna elements. Partial or reduced-dimensional channel state information (CSI) \cite{3DBeamforming_SPM2014} can be used to address this issue at the expense of channel estimation error.  Moreover, channel estimation error increases with the link distance, resulting from the signal-to-noise ratio deterioration and nonnegligible quantization errors \cite{BeamformingError_CL}. Furthermore, outdated CSI, multiuser interferences and hareware impairments can also cause channel estimation error \cite{OutdatedChannelEstError,PhaseShifterImpairment_SPL}. All these factors lead to pointing error and loss of beamforming gains.

Positioning information can be exploited to resolve the practical difficulties of beamforming. This idea originated from the fact that the desired antenna phase shifts can be expressed as functions of receiver locations in line-of-sight cases \cite{LocationBeamforming,WangW_IRS_BF_Positioning2021}, and {{positioning outputs can provide prior knowledge on conventional beamforming to help alleviate the complexity \cite{MultiLevelBeamforming_CL,PA_beamforming_Access,Bidirectional_Positioning_BeamformingTCOM}.}} It was shown that the positioning information can be used to design beamforming in 5G networks \cite{PositioningRailway5G_Mag}, and a subsequent work analyzed a beamformed radio link capacity with positioning errors \cite{BeamformingPositioningError_TVT}. However, the resulting capacity expression contains multifold integrals. An elegant performance expression is crucial especially when the system needs instantaneous parameter adjustment, which is the case for the positioning-assisted beamforming. In such systems, there exists a tradeoff between beamwidth and directivity \cite{BeamwidthTradeoff}, but the existing analytical results were based on the dichotomy of the beam's coverage and the receiver's movement is one-dimensional. A recent work showed that the estimated channel parameters could be refined based on the positioning information, but did not provide detailed performance analysis on the positioning-assisted approach \cite{WangW_IRS_BF_Positioning2021}.

In this work, {{we analyze the outage performance of positioning-assisted beamforming and optimize the beamwidth to suppress the outage probability. Closed-form outage probability bounds are derived and their asymptotic tightness is verified. Besides, the optimal beamwidth is expressed as a closed-form function of the link distance and the transmit power.}} The new analytical tool reveals insights into the design of reliable beamforming systems by exploiting positioning results, and lowers the computational complexity for parameter optimization. We show that the conventional wisdom does not necessarily hold for the narrowest beams to achieve the highest reliability.
\vspace{-10 pt}
\section{System Model}
\begin{figure}
  \centering
  \includegraphics[width=0.65\linewidth]{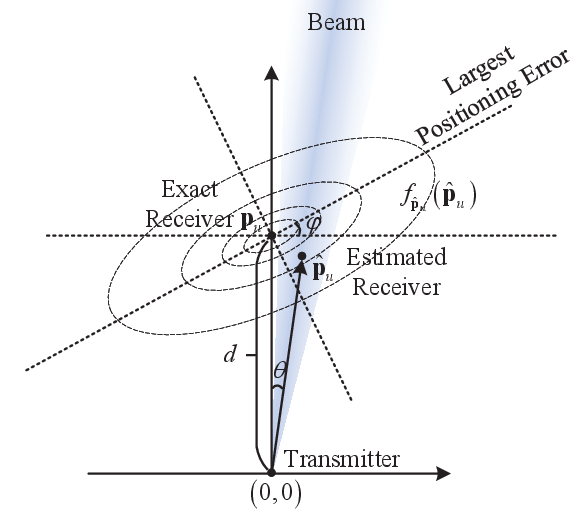}\\
  \caption{The beam direction and the exact and estimated receiver positions. The receiver is located at $\textbf{p}_u = (0,d)$, and the beam points to the estimated receiver position $\hat{ \textbf{p}}_u = (\hat x_u, \hat y_u)$. }\label{SystemModel}
\end{figure}

{{
\subsection{Positioning-Assisted Beamforming}
As shown in Fig. \ref{SystemModel}, we assume that the transmitter is located at $\left(0, 0\right)$ and the exact receiver position is ${\textbf{p}}_u=\left[0, d\right]^T$, where $d$ is the distance between the transmitter and the receiver. The estimated user position is $\hat{ \textbf{p}}_u = {\left[ {\hat x_u,\hat y_u} \right]^T}$. Before data transmission, the transmitter rotates the boresight direction of the beam towards the estimated user position $\hat{ \textbf{p}}_u$, where the pointing error angle is $\theta$. Therefore, we have $\theta=\textrm{atan2}(\hat y_u, \hat  x_u) \in (-\pi, \pi]$, where $\textrm{atan2}(\cdot,\cdot)$ is the four-quadrant inverse tangent function. Note that many detailed beamforming processes have been parameterized by $\theta$, including the generalization of the steering vector.
}}
\vspace{-5 pt}
\subsection{Positioning Error}
The joint probability density function of $\hat{ \textbf{p}}_u $ can be expressed as \cite{PA_beamforming_Access}
\begin{equation}\label{eq-1}
  {f_{\hat{ \textbf{p}}_u}}\left(\hat{ \textbf{p}}_u \right) = \frac{1}{{2\pi \sqrt {\det \left( {\textbf{R}} \right)} }}\exp \left( { - \frac{{{{\left( {\hat{ \textbf{p}}_u - {\textbf{p}}_u} \right)}^T}{{\textbf{R}}^{ - 1}}\left( {\hat{ \textbf{p}}_u - {\textbf{p}}_u} \right)}}{2}} \right)
\end{equation}
where ${\textbf{R}} = E\left[ {\left( {\hat{ \textbf{p}}_u- { \textbf{p}}_u} \right)\left( {\hat{ \textbf{p}}_u- { \textbf{p}}_u} \right)^T} \right]$ is the covariance matrix of the positioning error vector $\hat{ \textbf{p}}_u - { \textbf{p}}_u$, and $\det(\cdot)$ denotes the determinant. Applying eigenvalue decomposition, we can express $\textbf{R}$ as
\begin{equation}\label{eq-2}
{\textbf{R}} = \underbrace {\left[ {\begin{array}{*{20}{c}}
{\cos \varphi }&{ - \sin \varphi }\\
{\sin \varphi }&{\cos \varphi }
\end{array}} \right]}_{\textbf{G}}\underbrace {\left[ {\begin{array}{*{20}{c}}
{\sigma _1^2}&0\\
0&{\sigma _2^2}
\end{array}} \right]}_{\bf{\Lambda }}\\
  \underbrace {\left[ {\begin{array}{*{20}{c}}
{\cos \varphi }&{\sin \varphi }\\
{ - \sin \varphi }&{\cos \varphi }
\end{array}} \right]}_{{{\textbf{G}}^T}}
\end{equation}
where $\sigma_1>\sigma_2$, and $\varphi$ denotes the directional angle along which the larger positioning error occurs, which has variance $\sigma_1^2$, while $\pi/2-\varphi$ corresponds to the direction having the smaller positioning error with variance $\sigma_2^2$. This model complies with the conclusion that a Kalman filter outputs predictive Gaussian distributions when the prior observations follow a Gaussian distribution \cite{anderson1979optimalFiltering}.

\vspace{-3 pt}
\subsection{Channel Model}
{{Assuming the effective area of the receiver's antenna is $A_e$, and the transmitter antenna gain at the boresight direction is $G$, and the input power of the transmitter antenna is $P_{0}$, we can express the received power as\footnote{{We only focus on the pointing error caused fading to ensure a mathematically tractable channel model.}} \cite[Chap. 1]{ModernAntennaDesign2005}
\begin{equation}\label{eq-3}
{P_r} = {P_{\max }}A_{e}{G_\theta }{G_d}
\end{equation}
where $P_{\max }=P_{0} G$ ;  ${G_\theta }$ and ${G_d}$ denote, respectively, the normalized transmit antenna gain and the free-space propagation loss, which can be calculated through
\begin{equation}\label{eq-4}
{G_d} = 1/{{{{\left( {4\pi d} \right)}^2}}}
\end{equation}}}
and \cite[eq. (2)]{3DBeamforming_SPM2014}
\begin{equation}\label{eq-5}
{G_\theta } = \max \left\{ {{{10}^{ - 1.2\frac{{{\theta ^2}}}{{\theta _{3dB}^2}}}},{a_m}} \right\}
\end{equation}
where $\theta \in (-\pi,\pi]$ denotes the angle of the receiver direction from the boresight direction and $2\theta _{3dB}$ indicates the 3-dB beamwidth of the antenna radiation pattern, and {{$a_m<1$}} equals the side-lobe level attenuation of the antenna pattern. {{ The parameters $\theta_{3dB}$ and $a_m$ can model the beams generated by a variety of antennas.}}

 The outage probability of the positioning-assisted beamforming can be expressed as
\begin{equation}\label{eq-6}
\begin{split}
&{P_{out}}\left( {{\gamma _{th}}} \right) = \Pr \left( {{P_r} \le {\gamma _{th}}} \right)= \Pr \left( {{P_{\max }}A_{e}{G_\theta }{G_d} \le {\gamma _{th}}} \right)\\
 &= \Pr \left( {\max \left\{ {{{10}^{ - \frac{{1.2}}{{\theta _{3dB}^2}}{{ \textrm{atan2} }^2}\left( {{{{{\hat y}_u}}},{{{{\hat x}_u}}}} \right)}},{a_m}} \right\}\frac{{{A_e}}}{{{{\left( {4\pi d} \right)}^2}}} \le \frac{{{\gamma _{th}}}}{{{P_{\max }}}}} \right)\\
 &= \left\{ \begin{array}{l}
\Pr \left( {\left| {{{\hat x}_u}} \right| \ge k{{\hat y}_u}} \right),\frac{{{P_{\max }}{A_e}}}{{{{\left( {4\pi d} \right)}^2}}} > {\gamma _{th}} > \frac{{{P_{\max }}{a_m}{A_e}}}{{{{\left( {4\pi d} \right)}^2}}}\\
0,\ \ \ \ \ \ \ \ \ \ \ \ \ \ \ \ \ \frac{{{P_{\max }}{a_m}{A_e}}}{{{{\left( {4\pi d} \right)}^2}}} \ge {\gamma _{th}}\\
1,\ \ \ \ \ \ \ \ \ \ \ \ \ \ \ \ \ \frac{{{P_{\max }}{A_e}}}{{{{\left( {4\pi d} \right)}^2}}} \le {\gamma _{th}}
\end{array} \right.
\end{split}
\end{equation}
where $k = \tan \sqrt {\frac{{\theta _{3dB}^2}}{{1.2}}\lg \left( {\frac{{{A_e}{P_{\max }}}}{{{{\left( {4\pi d} \right)}^2}{\gamma _{th}}}}} \right)}>0 $; ${P_{out}}\left( {{\gamma _{th}}} \right)=0$ corresponds to the case where the received lobe power surpasses the threshold $\gamma_{th}$ and ${P_{out}}\left( {{\gamma _{th}}} \right)=1$ corresponds to the case where the received power is below the threshold even with perfect pointing.

\begin{figure}
  \centering
  \includegraphics[width=0.65\linewidth]{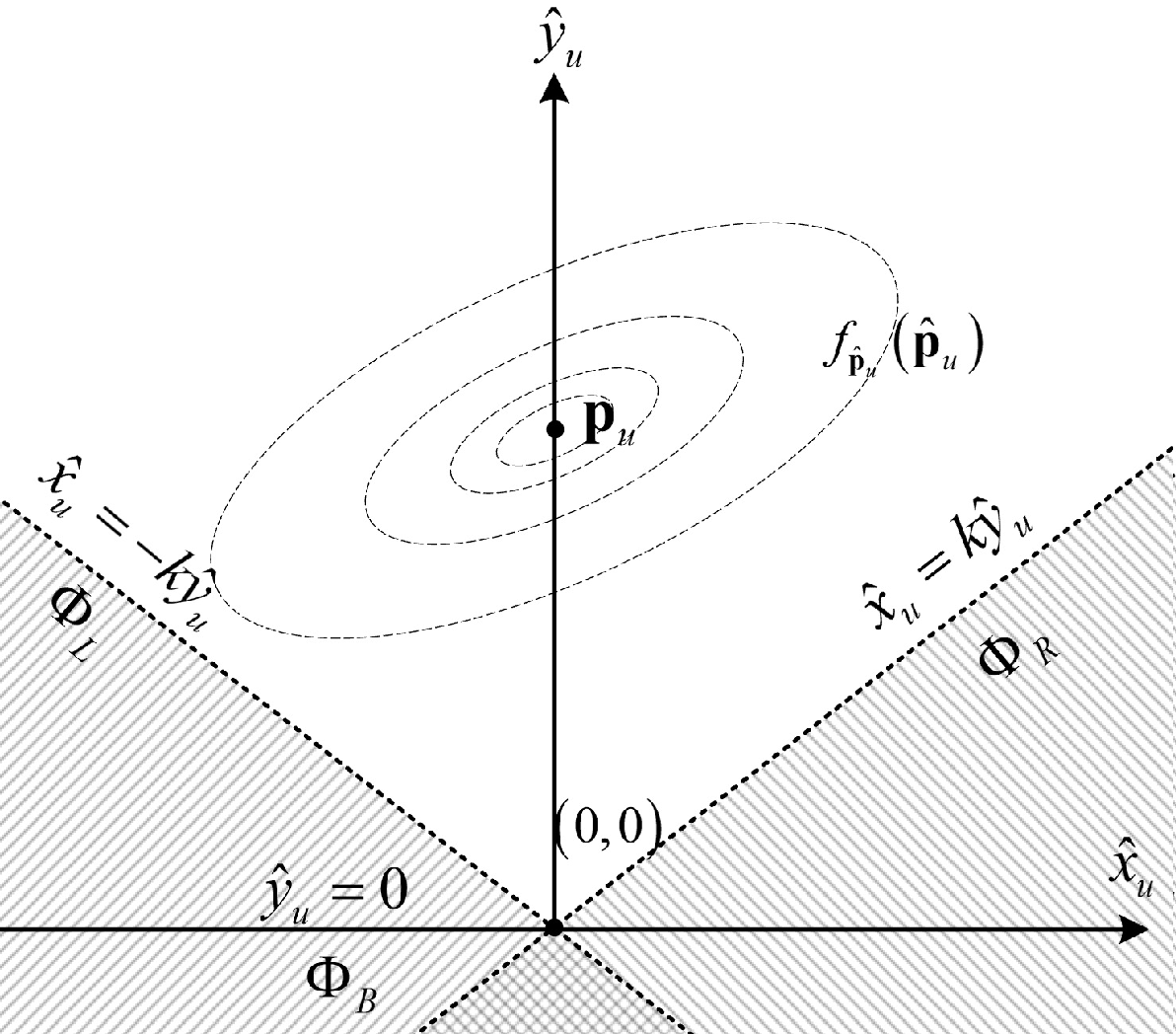}\\
  \caption{The beam direction and the exact and estimated receiver positions. The receiver is located at $\textbf{p}_u = (0,d)$, and the beam points to the estimated receiver position $\hat{ \textbf{p}}_u = (\hat x_u, \hat y_u)$. }\label{IntegralRegion}
\end{figure}

\section{Outage Probability}
\subsection{Outage Probability Bounds}
As shown in \eqref{eq-1} and \eqref{eq-6}, when $\frac{{{P_{\max }}{A_e}}}{{{{\left( {4\pi d} \right)}^2}}} > {\gamma _{th}} > \frac{{{P_{\max }}{a_m}{A_e}}}{{{{\left( {4\pi d} \right)}^2}}}$, the outage probability can be expressed
\begin{equation}\label{eq-7}
{P_{out}}\left( {{\gamma _{th}}} \right) = \iint\limits_{\left| {{{\hat x}_u}} \right| \ge k{{\hat y}_u}} {f_{\hat{ \textbf{p}}_u}}\left(\hat{ \textbf{p}}_u \right) d{\hat{ \textbf{p}}_u}
\end{equation}
where $d\hat{ \textbf{p}}_u:=d \hat x_u d \hat y_u$, and the integral region can be defined as
\begin{equation}\label{eq-8}
  {\Phi _{out}} = \left\{ {\left. {\left( {{{\hat x}_u},{{\hat y}_u}} \right)} \right|\left| {{{\hat x}_u}} \right| \ge k{{\hat y}_u}} \right\}.
\end{equation}
It can be verified that
\begin{equation}\label{eq-9}
  {\Phi _{out}} = {\Phi _L} \cup {\Phi _R}
\end{equation}
where
\begin{equation}\label{eq-10}
\begin{split}
{\Phi _R} &= \left\{ {\left. {\left( {{{\hat x}_u},{{\hat y}_u}} \right)} \right|{{\hat x}_u} \ge k{{\hat y}_u}} \right\},\\
{\Phi _L} &= \left\{ {\left. {\left( {{{\hat x}_u},{{\hat y}_u}} \right)} \right|{{\hat x}_u} \le  - k{{\hat y}_u}} \right\}
\end{split}
\end{equation}
which are shown in Fig. \ref{IntegralRegion}. Therefore, we can modify the integral region in \eqref{eq-7} analogously to \cite{RightTail_SLN_GC} and obtain
\begin{equation}\label{eq-11}
{P_{out}}\left( {{\gamma _{th}}} \right) < \underbrace {\iint\limits_{{{\hat x}_u} \ge k{{\hat y}_u}} {{f_{\hat{ \textbf{p}}_u}}\left( \hat{ \textbf{p}}_u \right)d{\hat{ \textbf{p}}_u}} }_{{I_R}} + \underbrace {\iint\limits_{{{\hat x}_u} \le  - k{{\hat y}_u}} {{f_{\hat{ \textbf{p}}_u}}\left( {{\hat{ \textbf{p}}_u}} \right)d\hat{ \textbf{p}}_u} }_{{I_L}}.
\end{equation}
and after complicated simplification in Appendix \ref{Integral_Simplify}, we have
\begin{equation}\label{eq-12}
{I_R} = Q\left( { - \frac{{\left[ {1, - k} \right]{{\textbf{p}}_u}}}{{\left\| {\left[ {1, - k} \right]\sqrt {\textbf{R}} } \right\|}}} \right)
\end{equation}
where $Q(x)=1/\sqrt{2\pi}\int_{x}^{\infty}\exp\{-t^2/2\}dt$ is the Gaussian $Q$-function and
\begin{equation}\label{eq-13}
{I_L} = Q\left( {\frac{{\left[ {1,k} \right]{{\textbf{p}}_u}}}{{\left\| {\left[ {1,k} \right]\sqrt {\textbf{R}} } \right\|}}} \right).
\end{equation}

According to Fig. \ref{IntegralRegion}, we can express the outage probability as
\begin{equation}\label{eq-14}
\begin{split}
{{\mathop{\rm P}\nolimits} _{out}}\left( {{\gamma _{th}}} \right) =
{I_R} + {I_L} - \int\limits_{{\Phi _R} \cap {\Phi _R}} {{f_{\hat{ \textbf{p}}_u}}\left( \hat{ \textbf{p}}_u \right)d} \hat{ \textbf{p}}_u
\end{split}
\end{equation}
where the integral region is bounded as
\begin{equation}\label{eq-15}
  {\Phi _L} \cap {\Phi _R} \subset \left\{ {\left. {\left( {{{\hat x}_u},{{\hat y}_u}} \right)} \right|{{\hat y}_u} \le 0} \right\}: = {\Phi _B}
\end{equation}
because
\begin{equation}\label{eq-16}
\left\{ \begin{array}{l}
{{\hat x}_u} \ge k{{\hat y}_u}\\
{{\hat x}_u} \le  - k{{\hat y}_u}
\end{array} \right. \Rightarrow \left\{ \begin{array}{l}
{{\hat x}_u} \ge k{{\hat y}_u}\\
 - {{\hat x}_u} \ge k{{\hat y}_u}
\end{array} \right. \Rightarrow  {{\hat y}_u} \le 0.
\end{equation}
Using the bounding region in \eqref{eq-15} to replace the integral region in \eqref{eq-14}, we obtain
\begin{equation}\label{eq-17}
  {{\mathop{\rm P}\nolimits} _{out}}\left( {{\gamma _{th}}} \right) > {I_R} + {I_L} - \underbrace {\int\limits_{{{\hat y}_u} \le 0} {{f_{\hat{ \textbf{p}}_u}}\left( \hat{ \textbf{p}}_u \right)d} {\hat{ \textbf{p}}_u}}_{{I_B}}
\end{equation}
where $I_B$ can be simplified according to Appendix \ref{Integral_Simplify} as
\begin{equation}\label{eq-18}
  I_B = Q\left( {\frac{{\left[ {0,1} \right]{{\textbf{p}}_u}}}{{\left\| {\sqrt {\textbf{R}} {{\left[ {0,1} \right]}^T}} \right\|}}} \right).
\end{equation}
Substituting \eqref{eq-12} and \eqref{eq-13} into \eqref{eq-11}, and \eqref{eq-12}, \eqref{eq-13}, \eqref{eq-18} into \eqref{eq-17}, we obtain the upper and lower bounds as
\begin{equation}\label{eq-19}
\begin{split}
&\underbrace {Q\left( { - \frac{{\left[ {1, - k} \right]{{\textbf{p}}_u}}}{{\left\| {\left[ {1, - k} \right]\sqrt {\textbf{R}} } \right\|}}} \right)}_{I_R} + \underbrace {Q\left( {\frac{{\left[ {1,k} \right]{{\textbf{p}}_u}}}{{\left\| {\left[ {1,k} \right]\sqrt {\textbf{R}} } \right\|}}} \right)}_{I_L} > {{\mathop{\rm P}\nolimits} _{out}}\left( {{\gamma _{th}}} \right)\\
 &> \underbrace {Q\left( { - \frac{{\left[ {1, - k} \right]{{\textbf{p}}_u}}}{{\left\| {\left[ {1, - k} \right]\sqrt {\textbf{R}} } \right\|}}} \right)}_{I_R} + \underbrace {Q\left( {\frac{{\left[ {1,k} \right]{{\textbf{p}}_u}}}{{\left\| {\left[ {1,k} \right]\sqrt {\textbf{R}} } \right\|}}} \right)}_{I_L}\\
 &- \underbrace {Q\left( {\frac{{\left[ {0,1} \right]{{\textbf{p}}_u}}}{{\left\| {\left[ {0,1} \right]\sqrt {\textbf{R}} } \right\|}}} \right)}_{I_B}.
\end{split}
\end{equation}

\subsection{Convergence of the Bounds}
To prove the tightness of the bounds, we need to develop an equality and two inequalities. Reforming \eqref{eq-12}, \eqref{eq-13} and \eqref{eq-18}, we obtain
\begin{equation}\label{eq-20}
\begin{split}
{I_R} &= Q\left( {  {d}/{{\left\| {\left[ { - 1/k,1} \right]\sqrt {\textbf{R}} } \right\|}}} \right),\\
{I_L} &= Q\left( {{d}/{{\left\| {\left[ {1/k,1} \right]\sqrt {\textbf{R}} } \right\|}}} \right),\\
{I_B} &= Q\left( {{d}/{{\left\| {\left[ {0,1} \right]\sqrt {\textbf{R}} } \right\|}}} \right)
\end{split}
\end{equation}
where the denominators can be compared as
\begin{equation}\label{eq-21}
\begin{split}
&\left\| {\left( { - \frac{1}{k},1} \right)\sqrt {\textbf{R}} } \right\|{\rm{ = }}\left\| { - \frac{1}{k}{{\left( {\sqrt {\textbf{R}} } \right)}_1}{\rm{ + }}{{\left( {\sqrt {\textbf{R}} } \right)}_2}} \right\|\\
 &= \sqrt {{{\left\| {\frac{1}{k}{{\left( {\sqrt {\textbf{R}} } \right)}_1}} \right\|}^2} + {{\left\| {{{\left( {\sqrt {\textbf{R}} } \right)}_2}} \right\|}^2} - \frac{2}{k}{{\left( {\sqrt {\textbf{R}} } \right)}_1} \cdot {{\left( {\sqrt {\textbf{R}} } \right)}_2}} \\
 &> \left\| {{{\left( {\sqrt {\textbf{R}} } \right)}_2}} \right\| = \left\| {\left[ {0,1} \right]\sqrt {\textbf{R}} } \right\|
\end{split}
\end{equation}
when ${( {\sqrt {\textbf{R}} } )_1} \cdot {( {\sqrt {\textbf{R}} } )_2} < 0$ where $( {\sqrt {\textbf{R}} } )_l$ denotes the $l$th row vector of $\sqrt{\textbf{R}}$, and
\begin{equation}\label{eq-22}
\begin{split}
&\left\| {\left( {\frac{1}{k},1} \right)\sqrt {\textbf{R}} } \right\|{\rm{ = }}\left\| {\frac{1}{k}{{\left( {\sqrt {\textbf{R}} } \right)}_1}{\rm{ + }}{{\left( {\sqrt {\textbf{R}} } \right)}_2}} \right\|\\
 &= \sqrt {{{\left\| {\frac{1}{k}{{\left( {\sqrt {\textbf{R}} } \right)}_1}} \right\|}^2} + {{\left\| {{{\left( {\sqrt {\textbf{R}} } \right)}_2}} \right\|}^2} + \frac{2}{k}{{\left( {\sqrt {\textbf{R}} } \right)}_1} \cdot {{\left( {\sqrt {\textbf{R}} } \right)}_2}} \\
 &> \left\| {{{\left( {\sqrt {\textbf{R}} } \right)}_2}} \right\| = \left\| {\left[ {0,1} \right]\sqrt {\textbf{R}} } \right\|
\end{split}
\end{equation}
when ${( {\sqrt {\textbf{R}} } )_1} \cdot {( {\sqrt {\textbf{R}} } )_2} \ge 0$.

The ratio of two Gaussian $Q$-functions satisfies
\begin{equation}\label{eq-23}
\begin{split}
&\mathop {\lim }\limits_{t \to \infty } \frac{{Q\left( {\alpha t} \right)}}{{Q\left( t \right)}}= \frac{{\int\limits_{\alpha t}^\infty  {\frac{1}{{\sqrt {2\pi } }}\exp \left( { - \frac{{{x^2}}}{2}} \right)dx} }}{{\int\limits_t^\infty  {\frac{1}{{\sqrt {2\pi } }}\exp \left( { - \frac{{{x^2}}}{2}} \right)dx} }}\\
 &= \mathop {\lim }\limits_{t \to \infty } \frac{{\frac{1}{{\sqrt {2\pi } }}\exp \left( { - \frac{{{{\left( {\alpha t} \right)}^2}}}{2}} \right)\alpha }}{{\frac{1}{{\sqrt {2\pi } }}\exp \left( { - \frac{{{t^2}}}{2}} \right)}}=\left\{ \begin{array}{l}
\infty ,0<\alpha  < 1\\
0,\alpha  > 1
\end{array} \right.
\end{split}
\end{equation}
where  the second equality is based on the L'Hospital's rule.

The bounds in \eqref{eq-19} are asymptotically tight in two cases.

\subsubsection{Case 1: $d \to \infty$} As shown in \eqref{eq-21} and \eqref{eq-22},  the denominator in $I_B$ in \eqref{eq-20} is never the biggest among the three denominators. Therefore, applying \eqref{eq-23} we can show
\begin{equation}\label{eq-24}
  \mathop {\lim }\limits_{d \to \infty } \frac{{{I_B}}}{{{I_L} + {I_R}}} = 0
\end{equation}
which leads to the conclusion that the bounds in  \eqref{eq-19} are asymptotically tight when $d \to \infty$.

\subsubsection{Case 2: ${{\rm{tr}}\left( {\textbf{\bf{R}}} \right)} \to 0$}
The covariance matrix $\textbf{R}$ can be normalized as ${{\textbf{R}}_N} = {{\textbf{R}}}/{{{\rm{tr}}\left( {\textbf{R}} \right)}}$, which can be assumed to be fixed when the relative positions of the receiver and the basestation are fixed, thus \eqref{eq-20} becomes
\begin{equation}\label{eq-25}
\begin{split}
{I_R} &= Q\left( {{{d/\sqrt {{\rm{tr}}\left( {\textbf{R}} \right)} }}/{{\left\| {\left[ { - 1/k,1} \right]\sqrt {{{\textbf{R}}_N}} } \right\|}}} \right),\\
{I_L} &= Q\left( {{{d/\sqrt {{\rm{tr}}\left( {\textbf{R}} \right)} }}/{{\left\| {\left[ {1/k,1} \right]\sqrt {{{\textbf{R}}_N}} } \right\|}}} \right),\\
{I_B} &= Q\left( {{{d/\sqrt {{\rm{tr}}\left( {\textbf{R}} \right)} }}/{{\left\| {\left[ {0,1} \right]\sqrt {{{\textbf{R}}_N}} } \right\|}}} \right).
\end{split}
\end{equation}
Applying \eqref{eq-21}-\eqref{eq-23} to compare the equations in \eqref{eq-25}, we have
\begin{equation}\label{eq-26}
  \mathop {\lim }\limits_{{\rm{tr}}\left( {\textbf{R}} \right) \to 0 } \frac{{{I_B}}}{{{I_L} + {I_R}}} = 0
\end{equation}
which leads to the conclusion that the bounds in \eqref{eq-19} are asymptotically tight when ${\rm{tr}}\left( {\textbf{R}} \right) \to 0 $, which corresponds to the cases with low positioning errors.

Case 1 describes the case when the transmitter and the receiver are sufficiently far away, and case 2 describes the case when the positioning system is sufficiently accurate.

\section{Optimization of the beamwidth}
The beamwidth $\theta_{3dB}$ can be adjusted at the transmitter to minimize the outage probability, and we assume the transmitted power $P_t$ to be fixed, which can be calculated as
\begin{equation}\label{eq-27}
\begin{split}
P_t=\int\limits_{ - {\pi} }^{{\pi}}  {{P_{\max }}{{10}^{ - 1.2\frac{{{\theta ^2}}}{{\theta _{3dB}^2}}}}d\theta } = {P_{\max }}\frac{{\sqrt \pi  erf \left( {{\pi} \sqrt {\frac{{1.2}}{{\theta _{3dB}^2}}\ln 10} } \right)}}{{\sqrt {\frac{{1.2}}{{\theta _{3dB}^2}}\ln 10} }}
\end{split}
\end{equation}
where $erf(x)=2/\sqrt{\pi} \int_0^x \exp(-t^2) dt$ is the Gaussian error function, and the side-lobe level is assumed to be negligible. {When $\theta_{3dB}<2.24$ rad, we have $erf \left( {{\pi} \sqrt {\frac{{1.2}}{{\theta _{3dB}^2}}\ln 10} } \right)>0.999 \approx 1$}, thus
\begin{equation}\label{eq-28}
\begin{split}
P_t  \approx {P_{\max }}\frac{{{\theta _{3dB}}\sqrt \pi  }}{{\sqrt {1.2\ln 10} }}
\end{split}
\end{equation}
and
\begin{equation}\label{eq-29}
\begin{split}
{P_{\max }} \approx {P_t}\frac{{\sqrt {1.2\ln 10} }}{{{\theta _{3dB}}\sqrt \pi  }}.
\end{split}
\end{equation}

To minimize the outage probability in \eqref{eq-7}, we need to minimize the integral region, which is equivalent to maximizing the factor $k$. The problem has the same maximizer with the following problem
\begin{equation}\label{eq-30}
 \mathop {\max }\limits_{{\theta _{3dB}}} \theta _{3dB}^2\lg \left( {\frac{{{A_e}{P_t}\sqrt {1.2\ln 10} }}{{{\theta _{3dB}}\sqrt \pi  {{\left( {4\pi d} \right)}^2}{\gamma _{th}}}}} \right)
\end{equation}
where $P_{\max}$ is replaced according to \eqref{eq-29} and we focus on the case ${\theta _{3dB}} \le \frac{{{A_e}{P_t}\sqrt {1.2\ln 10} }}{{\sqrt \pi  {{\left( {4\pi d} \right)}^2}{\gamma _{th}}}}$ so that $k>0$. Taking the derivative of \eqref{eq-30} in terms of $\theta _{3dB}$ and making it zero, we can obtain the maximizer of \eqref{eq-30} as
\begin{equation}\label{eq-31}
\theta _{3dB}^* = \frac{{{A_e}{P_t}\sqrt {1.2\ln 10} }}{{\sqrt \pi  {{\left( {4\pi d} \right)}^2}{\gamma _{th}}}}{10^{ - \frac{1}{{2\ln 10}}}}
\end{equation}
which results in the optimal $k$ as
\begin{equation}\label{eq-32}
{k^*} = \tan \left( {\frac{{{A_e}{P_t}}}{{\sqrt {2\pi } {{\left( {4\pi d} \right)}^2}{\gamma _{th}}}}{{10}^{ - \frac{1}{{2\ln 10}}}}} \right).
\end{equation}
Equation \eqref{eq-31} indicates that the optimal beamwidth is a function of the transmit power $P_t$, the link distance $d$ and the outage threshold $\gamma_{th}$. When $d$ grows, the optimal beamwidth decreases to counter the pathloss; when $P_t$ grows, the optimal beamwidth increases to cover a larger angular range to counter the misalignment due to the positioning error.

\vspace{-5pt}
\section{Numerical Results}
\begin{figure}
  \centering
  \includegraphics[width=0.9\linewidth]{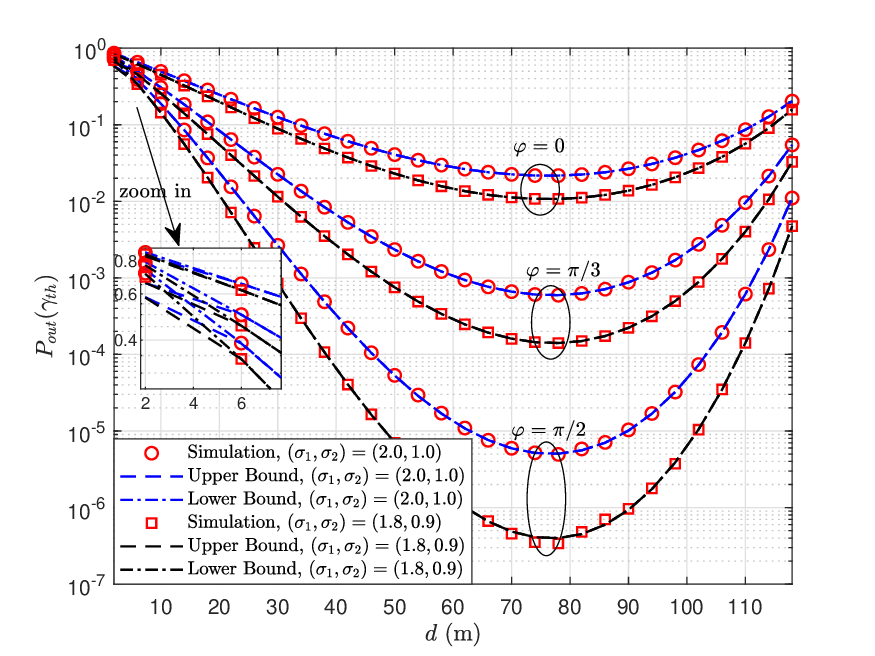}\\
  \caption{Outage probability as a function of the distance $d$. $P_{\max}=100$ W; $\gamma_{th}=10^{-7}$ W; $\theta_{3dB}=0.1$ rad; $a_m=10^{-4}$; {{ $A_e=5.0$ cm$^2$}}. }\label{distance}
\end{figure}
Figure \ref{distance} compares the outage probabilities with different positioning error, where $\theta_{3dB}$ is fixed to be $0.1$ rad ($5.73^\circ $). {{An antenna with the certain parameters $\theta_{3dB}$ and $a_m$ can be designed with the technique in  \cite[Chap. 3]{ModernAntennaDesign2005}.}} It is shown that the bounds in \eqref{eq-19} are sufficiently tight at the practical link distances. It can also be observed that the outage probability drops within the region $d\in(0, 75]$, and then rises in $d \in (75,120]$. An explanation is that when the receiver is too close to the transmitter, the pointing error angle $\theta$ in Fig. \ref{SystemModel} randomly varies in a large region, and using a fixed beam is not able to cover such a large angular range, resulting in a high outage probability. By comparing the black and blue curves, we can observe that the higher positioning accuracy, quantified by $\sigma_1$ and $\sigma_2$, can reduce the outage probability. Besides, the largest positioning error direction in Fig. \ref{SystemModel}, quantified by $\varphi$, also has significant impact on the outage probability. Specifically, when $\varphi=\pi/2$, the outage probability is the lowest, this is because the error along the $\varphi=\pi/2$ direction influences less on the pointing error angle $\theta$. These results imply that the basestations should be selected based on the positioning error pattern characterized by $\textbf{R}$.

\begin{figure}
  \centering
  \includegraphics[width=0.9\linewidth]{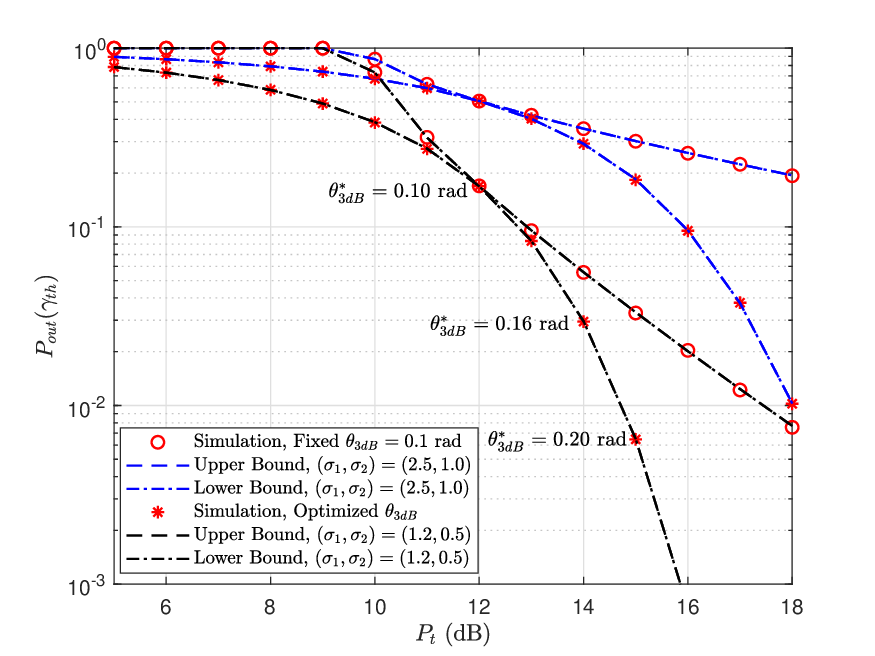}\\
  \caption{Outage probability as a function of the transmit power $P_t$. $d=30$ m; $\gamma_{th}=10^{-7}$ W; $a_m=10^{-2}$; $\varphi=\pi/4$; {{ $A_e=1.25$ cm$^2$}}. The fixed beamwidth systems have $\theta_{3dB}=0.1$ rad, and the optimized beamwidths are calculated by \eqref{eq-31}.}\label{Pt}
\end{figure}

Figure \ref{Pt} compares the outage probability with a fixed beamwidth and an optimized beamwidth. It can be observed again that the derived upper and lower bounds are sufficiently tight. Besides, the diversity order of the system is related to the positioning accuracy, which can be concluded by comparing the two curves with a fixed $\theta_{3dB}$ but different $(\sigma_1, \sigma_2)$. More importantly, by comparing the curves with the fixed and the optimized $\theta_{3dB}$'s calculated by \eqref{eq-31}, we can see that the optimized $\theta_{3dB}$ can lower the outage probability. More importantly, the outage probability drops dramatically without approaching a straight asymptote when $\theta_{3dB}$ is optimized. Another interesting observation is that $P_{out}(\gamma_{th})<1$ for all $P_t$'s when $\theta_{3dB}=\theta_{3dB}^*$, which is in contrast with the cases with fixed $\theta_{3dB}$'s. An explanation is that even though the transmit power is low, $\theta_{3dB}^*$ becomes small to maintain communication in a small angular range, and thus the outage event will not always happen. The explanation can be verified by the labels of $\theta_{3dB}$'s at different $P_t$'s, showing that $\theta_{3dB}^*$ grows with $P_t$. The observation agrees with the intuition that extra power can be allocated to the angular range that was previously uncovered.

\begin{figure}
  \centering
  \includegraphics[width=0.9\linewidth]{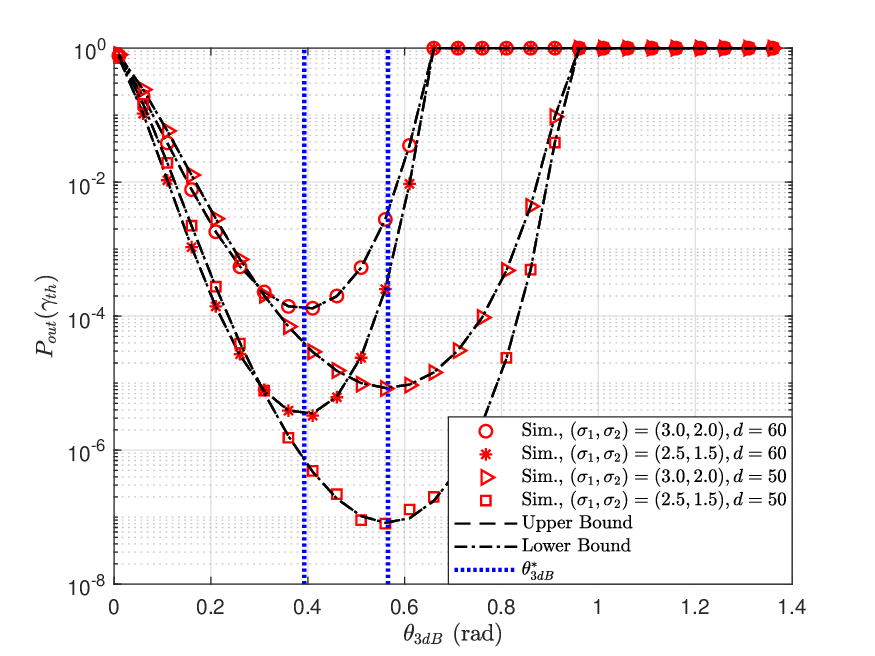}\\
  \caption{Outage probability as a function of the half-power beamwidth $\theta_{3dB}$. $\gamma_{th}=10^{-7}$ W; $a_m=10^{-2}$; $\varphi=\pi/4$; {{ $A_e=1.25$ cm$^2$}}; $P_t=24$ dB}\label{theta3dB}
\end{figure}
Figure \ref{theta3dB} shows how $\theta_{3dB}$ influences the outage probability. It is shown that the bounds in \eqref{eq-19} are tight for all $\theta_{3dB}$. More importantly, the figure shows four outage probability minima when $\theta_{3dB}$ changes, and they all agree with the analytical results in \eqref{eq-31}. It is shown that $\theta_{3dB}^*$ is related to $d$, implying that the beamwidth should be adjusted according to the receiver location; besides, the positioning accuracy, quantified by $(\sigma_1, \sigma_2)$, does not influence $\theta_{3dB}^*$. Another observation is that when $\theta_{3dB}$ is too large, we have $P_{out}(\gamma_{th})=1$, because the limited transmit power has been too divergent and unable to counter the pathloss. On the other hand, when the beam is too narrow, it is unlikely to cover the receiver when positioning error exists, as is predicted near $\theta_{3dB}=0$.

\vspace{-5pt}

\section{Conclusions}\label{conclusions}
We studied positioning-assisted beamforming scheme and its performance. Closed-form outage probability bounds were developed, whose asymptotic tightness was discussed. It was shown that the beamwidth should be optimized with respect to the link distance and transmit power, and the optimal beamwidth can be expressed in a closed form.

\begin{appendices}
\section{}\label{Integral_Simplify}
Changing the dummy variables $\hat {\textbf{p}}_u$ to ${\textbf{t}} = \sqrt {{{\textbf{R}}^{ - 1}}} \left( {\hat {\textbf{p}}_u- {{\textbf{p}}_u}} \right)$, we can express $I_R$ as
\begin{equation}\label{A-1}
{I_R} = \int\limits_{\left[ {1, - k} \right]\left( {\sqrt {\textbf{R}} {\textbf{t}} + {\textbf{p}}} \right) \ge 0} {\frac{{\exp \left( { - \frac{{{{\bf{t}}^T}{\textbf{t}}}}{2}} \right)\det \left( {\sqrt {\textbf{R}} } \right)}}{{2\pi \sqrt {\det \left( {\textbf{R}} \right)} }}d} {\textbf{t}}
\end{equation}
where the $\det ( {\sqrt {\textbf{R}} })$ is the Jacobian determinant. Then we construct an orthogonal matrix
\begin{equation}\label{A2}
{\textbf{W}}{\rm{ = }}\left[ {\underbrace {\frac{{\sqrt {\textbf{R}} {{\left[ {1, - k} \right]}^T}}}{{\left\| {\left[ {1, - k} \right]\sqrt {\textbf{R}} } \right\|}}}_{{{\textbf{w}}_1}},{{\textbf{w}}_2}} \right]
\end{equation}
where $\textbf{w}_2$ can be found through Schmidt orthogonalization so that ${\textbf{w}}_1^T{\textbf{w}}_2=0$ and $\|{\textbf{w}}_2\|=1$. Changing the dummy variables $\textbf{t}$ to ${\textbf{l}} = {{\textbf{W}}^T\textbf{t}}$, we have
\begin{equation}\label{A3}
\begin{split}
{I_R} &= \int\limits_{\left[ {1, - k} \right]\sqrt {\textbf{R}} {\textbf{Wl}} + \left[ {1, - k} \right]{{\textbf{p}}_u} \ge 0} {\frac{1}{2\pi }{\exp \left( { - \frac{{{{\textbf{l}}^T}{\textbf{l}}}}{2}} \right)}d} {\textbf{l}}\\
 &= Q\left( { - \frac{{\left[ {1, - k} \right]{{\textbf{p}}_u}}}{{\left\| {\left[ {1, - k} \right]\sqrt {\textbf{R}} } \right\|}}} \right)
\end{split}
\end{equation}
which proves \eqref{eq-12}. Following similar procedures we can prove \eqref{eq-13} and \eqref{eq-18}.

\end{appendices}

\bibliographystyle{IEEEtran}


\end{document}